\documentstyle[sprocl,epsfig]{article}

\def\Journal#1#2#3#4{{#1} {\bf #2}, #3 (#4)}

\def\NPB{{\em Nucl. Phys.} B}
\def\PLB{{\em Phys. Lett.}  B}

\def\PRD{{\em Phys. Rev.} D}

\def\be{\begin{equation}}
\def\ee{\end{equation}}
\def\bea{\begin{eqnarray}}
\def\eea{\end{eqnarray}}
\newcommand{\lm}[1]{\mbox{$\lambda_{#1}$}}
\newcommand{\m}[1]{\mbox{$m_{#1}^{2}$}}

\renewcommand{\H}[1]{\mbox{$H_{#1}$}}
\newcommand{\Hd}[1]{\mbox{$H_{#1}^{\dag}$}}

\begin{document}

GUTPA/99/2/01

\title{CP VIOLATION IN THE NEXT TO MINIMAL SUPERSYMMETRIC STANDARD
MODEL\footnote {To be published in Proceedings of the Conference on Strong and
Electroweak Matter, Copengagen, December 1998}}

\author{A.T. DAVIES, C.D. FROGGATT and A. USAI}

\address{Department of Physics and Astronomy, University of Glasgow,\\ Glasgow
G12 8QQ, Scotland,\\
E-mail: a.davies@physics.gla.ac.uk}

\maketitle\abstracts{ We consider spontaneous CP violation in the Next to
Minimal Supersymmetric Standard Model (NMSSM), without the usual $Z_3$ discrete
symmetry. CP violation can occur at tree level, raising a potential conflict
with the experimental bounds on the electric dipole moments of the electron and
neutron. One escape from this is to demand that the CP violating angles are
small, but we find that this entails an unacceptably light neutral Higgs.}

\section {Electroweak Baryogenesis} \label{EWB}

Two requirements of Electroweak Baryogenesis underlie this work. The generation
of baryon asymmetry from an initially symmetric state requires stronger CP
violation than is provided by the CKM matrix in the Standard Model, and the
Higgs sector is a possible source of this. Once baryon number has been created,
whether in the electroweak transition or earlier, there arises the problem of
preventing its wipeout by sphaleron transitions below the electroweak
transition, and this needs a strongly first order phase transition, which in
the Standard model entails an unwanted light Higgs. The Next to Minimal
Supersymmetric Standard Model (NMSSM) has some advantages over the Minimal
model (MSSM) in both these areas. As well as producing and preserving baryon
number, models have to avoid dangerous by-products, in particular a large
electric dipole moments arising from the CP violating phases. We find that a
light neutral Higgs also accompanies weak spontaneous CP violation.

Sphaleron transitions produce an irreconcileable conflict between electroweak
baryogenesis and the Standard Model. The MSSM can avoid this, at the expense of
a light stop and a neutral Higgs just above current experimental
reach~\cite{Carena,Del}. If these do not materialise, the NMSSM will deserve
increased consideration, as it has an extensive parameter space, much of which
is secure from baryon washout by sphalerons~\cite{Comelli,Dav1}.

CP violation (CPV) can arise explicitly, via complex couplings in the
Lagrangian, or spontaneously when the minima of a real potential occur at
complex vevs. The MSSM can incorporate explicit CPV by including complex phases
in the soft SUSY breaking terms involving the squarks and gauginos. It has
however no spontaneous CP violation (SCPV) at tree level, and generation by
radiative corrections entails an unacceptably light Higgs~\cite{Maekawa,Pom1}.
In contrast, in the  NMSSM  CPV can occur spontaneously even at tree level if
there is no additional discrete $Z_3$ symmetry~\cite{Pom2,Dav2}. The phases in
all these CPV models give rise to potentially large electric dipole moments.
The NMSSM with $Z_3$ has the attractive feature of permitting SCPV at finite
temperature where it can play a role in baryogenesis~\cite{Comelli}, but not at
zero temperature, thus avoiding problems with the electric dipole moments.
However, this model with the additional constraint of universal supersymmetry
breaking terms at the GUT scale has had difficulty fitting the observational
constraints on the Higgs spectrum~\cite{Haba}.

In order to disentangle the CP and Higgs spectrum properties we consider an
unconstrained NMSSM, without $Z_3$ and with general soft SUSY breaking
terms\footnote{Spontaneous breaking of a discrete symmetry raises a
cosmological domain wall problem.}.

\section{NMSSM} \label{NMSSM}

Our model is based on the superpotential
\begin {eqnarray}
W =  \lambda NH_1H_2 -\frac{k}{3} N^3 - r N + \mu H_1H_2 + W_{Fermion}
\end{eqnarray}
where $H_1$ and $H_2$ are the doublets of the MSSM and $N$ is a singlet.
We do not impose the common restriction $\mu = r =0$, which adds a discrete
$Z_3$ symmetry.  Nor do we require the soft SUSY-breaking terms to evolve
perturbatively from a universal high energy form. This allows two additional
$Z_3$-violating terms as well as more freedom in the  coupling constants.

At the electroweak scale the effective potential is~\cite{GunHab}
\begin{eqnarray}
\label{eq.vs0}
V_{0} & = & \frac{1}{2}\lm{1}(\Hd{1}\H{1})^2 + \frac{1}{2}\lm{2}(\Hd{2}\H{2})^2
  \nonumber  \\
  &  & + (\lm{3}+\lm{4})(\Hd{1}\H{1})(\Hd{2}\H{2}) - \lm{4}\left|
\Hd{1}\H{2}\right|^{2} \nonumber\\
  &  & + (\lm{5}\Hd{1}\H{1} +\lm{6}\Hd{2}\H{2})N^{\star}N+
(\lm{7}\H{1}\H{2}N^{\star 2}+h.c.)  \nonumber  \\
 &  &+  \lm{8}(N^{\star}N)^2+
\lambda \mu (N+h.c.)(\Hd{1}\H{1}+\Hd{2}\H{2})  \nonumber \\
 & & + \m{1}\Hd{1}\H{1}+\m{2}\Hd{2}\H{2} + \m{3}N^{\star}N -  m_4
(\H{1}\H{2}N+h.c.)
 \nonumber\\
 & & - \frac{1}{3}m_5(N^3+h.c.)+\frac{1}{2} \m{6}(\H{1}\H{2}+h.c.)+
\m{7}(N^2+h.c.)
\end{eqnarray}
where the quartic couplings $\lambda_i, i =1 \dots 8 $ at the electroweak scale
are related via renormalization group equations to the gauge couplings and the
$\lambda, k$ of the superpotential, assuming that the electroweak scale
$M_{Weak} < M_S$, the supersymmetry scale, taken to be 1 TeV.
$ m_i, i = 1 \dots 7 $, are taken as arbitrary parameters.
We consider  real coupling constants, so that the tree level potential is CP
conserving, but admit complex vevs for the  neutral fields, $\langle H_i^0
\rangle = v_i e^{i \theta_i} (i=1,2), \langle N \rangle=v_3 e^{i \theta_3}$,
giving
\begin{eqnarray} \label{eq.vsvac}
V_{0} &  = &  \frac{1}{2}(\lm{1}v_1^4  +\lm{2}v_2^4) +  (\lm{3}+\lm{4})v_1^2
v_2^2  +
(\lm{5}v_1^2 +\lm{6}v_2^2)v_3^2  \nonumber\\
&  & + 2\lm{7}v_1 v_2 v_3^2 cos(\theta_1 + \theta_2 - 2 \theta_3)+ \lm{8}v_3^4
+
2 \lambda\mu (v_1^2+v_2^2)v_3 cos(\theta_3)  \nonumber\\
 &  & + \m{1}v_1^2+\m{2}v_2^2 + \m{3} v_3^2-2m_4 v_1 v_2 v_3 cos(\theta_1 +
\theta_2 + \theta_3)
 \nonumber\\
 & & - \frac{2}{3} m_5 v_3^3 cos(3 \theta_3)+ 2\m{6}v_1 v_2cos(\theta_1 +
\theta_2) +2\m{7} v_3^2 cos(2 \theta_3)
\end{eqnarray}
where, without loss of generality,  $\theta_2 = 0$.
We trade some $m_i$ for chosen vevs $v_0 = \sqrt{v_1^2+v_2^2} = 174 \mbox{
GeV}$, $\tan \beta \equiv v_2/v_1$, \mbox{ $R \equiv v_3/v_0 $}, $\theta_1,
\theta_3$, as well as the tree level charged Higgs mass $M_{H^+}$, leaving
$m_5$ and $\mu$ free.

Sets of parameters are chosen which satisfy the conditions for a stationary
value of the potential, and then numerical searches are performed to ensure
that this is the global minimum.

\section{Spontaneous CP violation and the Higgs spectrum} \label{MHiggs}
The NMSSM with $Z_3$ does not allow SCPV~\cite{Romao}. If SUSY is broken by
radiative corrections to the quartic potential SCPV is possible but is
accompanied by a light scalar~\cite{Babu}. As noticed by Pomarol~\cite{Pom2}
inclusion of general soft breaking terms, namely those in $m_6^2$ and $m_7^2$
above, does allow SCPV. Large neutron and electron dipole moments can be
suppressed by three mechanisms, alone or in combination: (i) large squark and
gaugino masses (few TeV); (ii) small phases, $ O(0.01)$~\cite{Pom2}; or (iii)
cancellations between the graphs contributing. This last possibility is more
generic in the constrained MSSM than was hitherto realised~\cite{IbNath}.

We find that consideration of the Higgs spectrum disfavours the small phase
option in the SCPV case.

The scalar mass matrix gives rise to 1 charged and 5 neutral particles.  An
acceptable mass spectrum can readily be obtained~\cite{Dav2}. For example, the
parameters in Table~\ref{Tab:spectrum}, with $\lambda = k = 0.5,~M_S = 1
\mbox{~TeV}$ give neutral masses 89 to 318 GeV.
\begin{table}[t]
\caption{Example of parameters giving spontaneous CP violation and large Higgs
masses. \label{Tab:spectrum}}
\vspace{0.2cm}
\begin{center}
\footnotesize
\begin{tabular} {|c|c|c|c|c|c|c|} \hline
$\mbox{tan}\beta$ & R  ($\equiv  v_3/v_0$) &  $\theta_1$  &  $\theta_3$ &
$M_H^+$ & $m_5$ & $\mu $\\   \hline
2.0 & 2.0 & 1.20 $ \pi$ & 0.65$ \pi$ & 250 GeV & 60 GeV & -20 GeV\\ \hline
\end{tabular}
\end{center}
\end{table}
This example corresponds to large angles in the vevs, and indeed such a
spectrum arises for a wide range of parameters in the potential. Fig.~\ref{Fig:
theta 1.0} shows the upper bound on the lightest neutral when we require SCPV
and allow large angles $\theta \approx 1 $ radian.  $\theta_1$ was fixed at
$\theta_1 = 1$ and $\theta_3$ increased in steps from $0.5~\theta_1$ to
$2.0~\theta_1$.  100,000 sets of  other parameters in the potential were
randomly chosen, in the ranges $2< \tan \beta < 3$, and, in GeV, $10 \le v_3
\le 500$, $-500 \le \mu \le +500$, $0 \le m_5 \le 500$, $200 \le M_{H^+} \le
800$.  When both angles are large the lightest neutral can have a mass $m_h >
80 \mbox{ GeV}$. The mass of the lightest Higgs is always $< 122 \mbox{ GeV}$
which is not far above current experimental reach at LEP, but it can contain a
significant admixture of the singlet $N$ field, which reduces its experimental
visibility~\cite{Haba}.

\begin{figure}[h]
\unitlength1cm
\begin{minipage}[t]{5.5cm}
\begin{picture}(5.5,5.5)
\epsfig{file=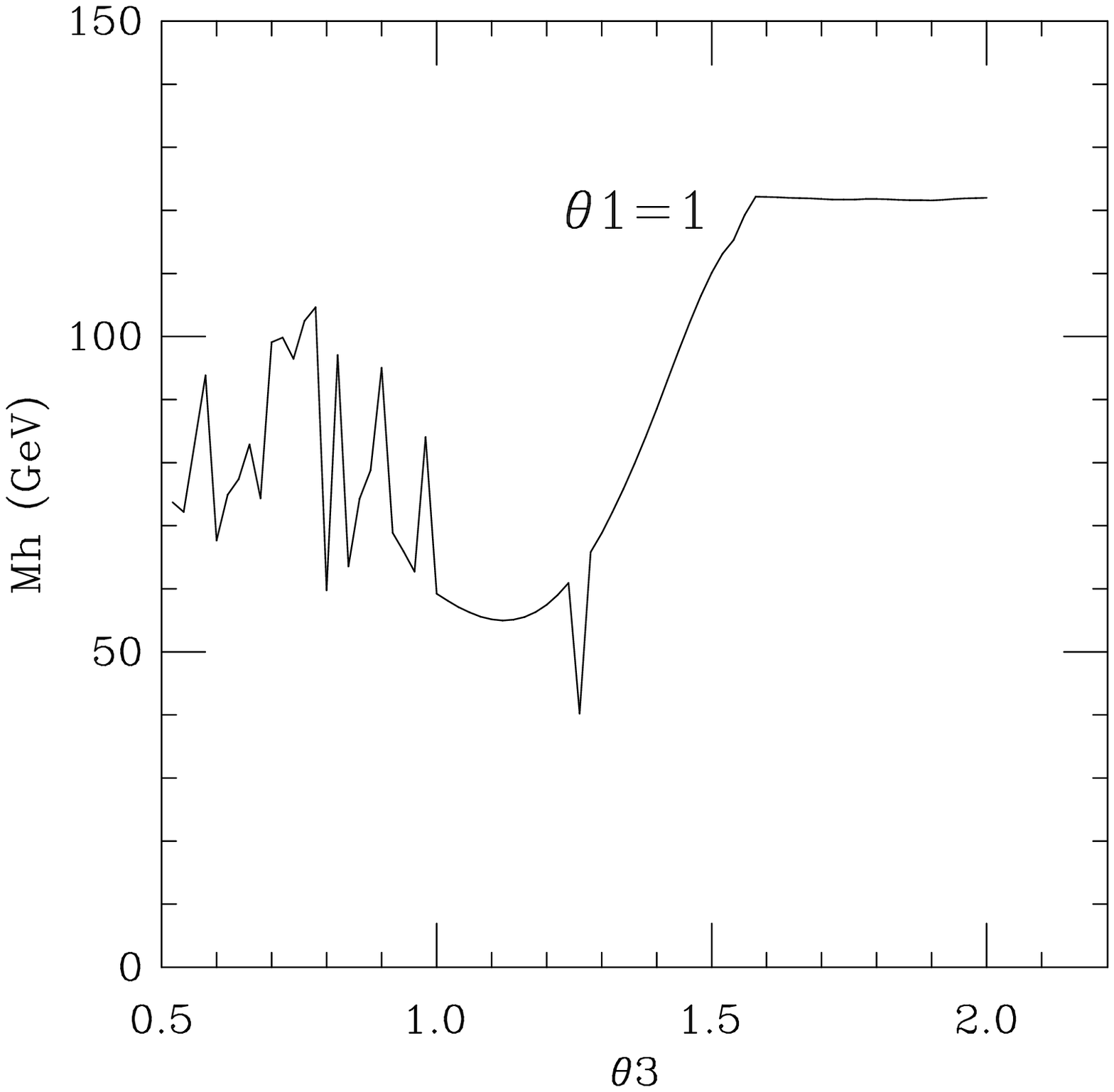, width=5.5cm}
\end{picture}\par
\caption{Large phases. Upper bound on the lightest neutral Higgs mass.
\label{Fig: theta 1.0}}
\end{minipage}
\hfill
\begin{minipage}[t]{5.5cm}
\begin{picture}(5.5,5.5)
\epsfig{file=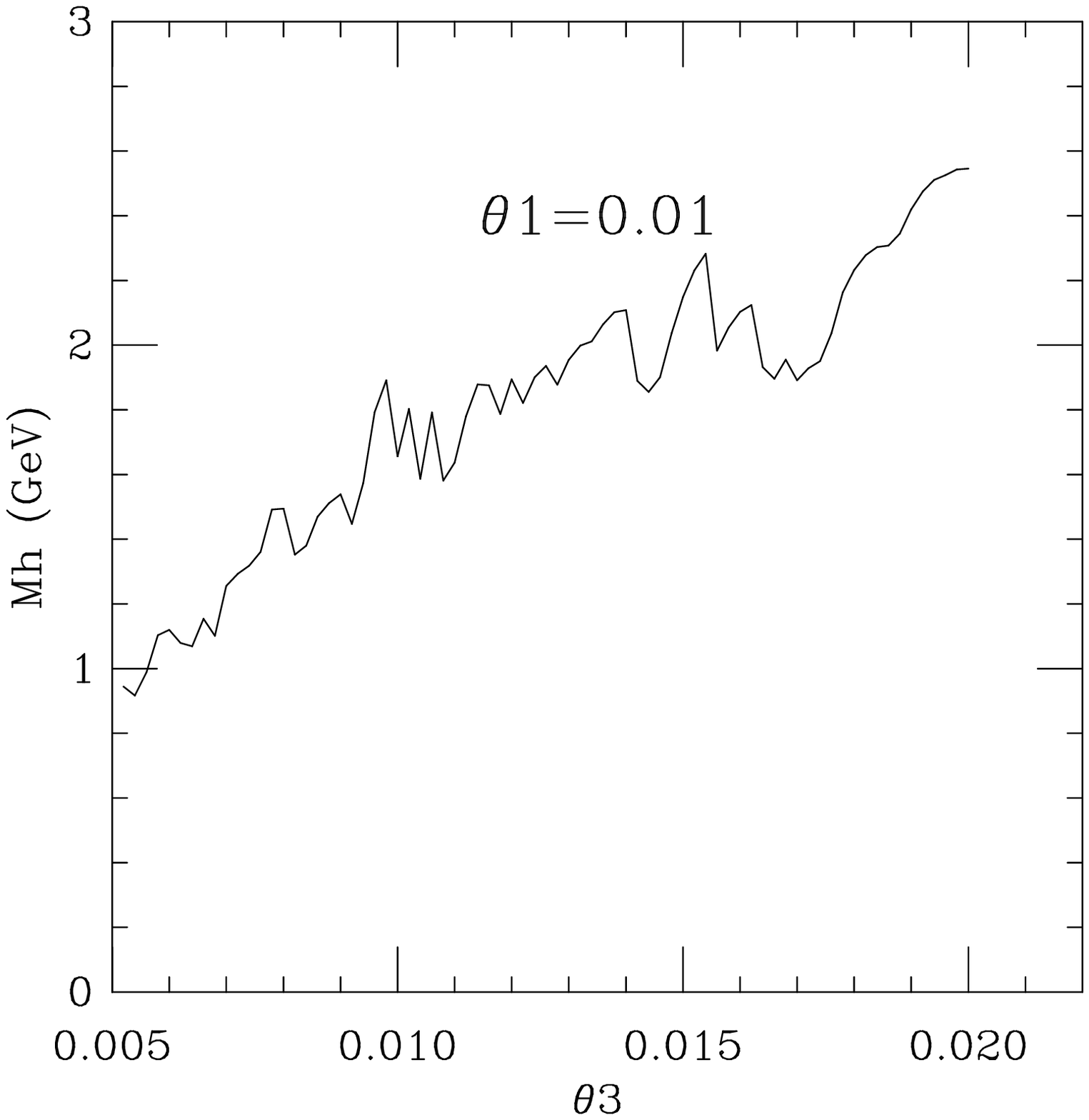, width=5.5cm}
\end{picture}\par
\caption{Small phases. Upper bound on the lightest neutral Higgs mass.
\label{Fig: theta 0.01}}
\end{minipage}
\end{figure}

Similar scans with small CP violating phases exhibit a neutral light scalar.
Phases $\theta \approx 0.1$ radians give a mass $m_h < 30$ GeV, and, as shown
in Fig.~\ref{Fig: theta 0.01}, $\theta_1 = 0.01,\mbox{~and} ~ 005 < \theta_3
\le 0.02$  give $m_h < 3$ GeV. Thus we can probably exclude possibility (ii)
above.

\section{Discussion} \label{Disc}
 The result that weak  spontaneous CP breaking implies a light Higgs is quite
general, and may be understood by an argument similar to that used by Georgi
and Pais~\cite{Georgi} in proving a theorem on the conditions under which
radiative corrections can trigger SCPV. If CP is weakly broken, the potential
has two nearby minima at

\begin{equation}
\underline{\epsilon}_1 =  (v_1,v_2,v_3, v_1 \theta_1, v_2 \theta_2, v_3
\theta_3),
\end{equation}
\begin{equation}
\underline{\epsilon}_2= (v_1,v_2,v_3,-v_1 \theta_1,-v_2 \theta_2,-v_3
\theta_3).
\end{equation} Performing a  Taylor expansion
\begin{eqnarray}
(\epsilon_2 - \epsilon_1)_j \frac{\partial^2 V}{\partial \phi_j \partial
\phi_i}\Big|_{\underline{\epsilon}_1} \approx
\frac{\partial V}{\partial \phi_i} \Big|_{\underline{\epsilon}_2} -
\frac{\partial V}{\partial \phi_i} \Big|_{\underline{\epsilon}_1}
= 0 - 0,
\end{eqnarray} so the mass squared matrix must be singular. To zeroth order
there is a zero mass particle, with eigenvector along the direction in the
6-dimensional neutral Higgs space joining the two CP violating minima. If
$\theta_i \neq 0$, the neutral matrix does not decouple into sectors with CP =
+1 and -1, but it does so approximately as the off diagonal blocks of the
matrix are proportional to the small angles $\theta$. This light particle is
thus in the nearly CP odd sector. Depending on the parameters in the potential
this particle can be a varying admixture of singlet and doublet fields, so it
may be difficult to detect.

We conclude that SCPV with small phases is disfavoured.  It can occur with
large phases, but then heavy squarks or cancellations are required to suppress
electric dipole moments.

\section*{References}

%


\begin{thebibliography}{99}
\bibitem{Carena} M. Carena, M. Quiros and C.E.M. Wagner,
\Journal{\PLB}{380}{81}{1996}.
\bibitem{Del} D. Delepine, J.M. Gerard, R. Gonzalez Filipe and J. Weyers,
\Journal{\PLB}{386}{183}{1996}.
\bibitem{Comelli} D. Comelli, M. Pietroni and A. Riotto,
\Journal{\PRD}{50}{7703}{1994}.
\bibitem{Dav1} A.T. Davies, C.D. Froggatt, G. Jenkins and R.G. Moorhouse,
\Journal{\PLB}{372}{88}{1996}.
\bibitem{Maekawa} N. Maekawa, \Journal{\PLB}{282}{392}{1992}.
\bibitem{Pom1} A. Pomarol, \Journal{\PLB}{287}{331}{1992}.
\bibitem{Pom2} A. Pomarol, \Journal{\PRD}{47}{273}{1993}.
\bibitem{Dav2} A.T. Davies, C.D. Froggatt and A. Usai,{\it~ Proc. of the
International Europhysics Conference on High Energy Physics}, Jerusalem (1997),
Eds. D. Lellouch, G. Mikenberg and E. Rabinovici , Springer-Verlag (1998),p891.
\bibitem{Haba}N. Haba, M. Matsuda and M.
Tanimoto,~\Journal{\PRD}{54}{6928}{1996}.
\bibitem{GunHab} J. Gunion and H.E. Haber, \Journal{\NPB}{272}{1}{1986}.
\bibitem{Romao} J. C. Romao, \Journal {\PLB}{173}{309}{1986}.
\bibitem{Babu} K.S. Babu and S.M. Barr, \Journal{\PRD}{49}{R2156}{1994}.
\bibitem {IbNath} T. Ibrahim and P. Nath \Journal{\PRD}{57}{478}{1998}.
\bibitem{Georgi} H. Georgi and A. Pais, \Journal{\PRD}{10}{1246}{1974}.
\end{thebibliography}
\end{document}